\documentclass[letterpaper]{aa}
\usepackage{graphicx}

\def\mlstar{$(M/L)_*$}
\def\mstar{M$_*$}

\begin{document}

\title{The stellar masses of 25000 galaxies at $0.2 \le z \le 1.0$\\
estimated by the COMBO-17 survey}

\author{Andrea Borch \inst{1,5} \and Klaus Meisenheimer \inst{1}
                              \and Eric F.\ Bell \inst{1}
                              \and Hans-Walter Rix \inst{1}
                              \and Christian Wolf \inst{2}
                               \and Simon Dye \inst{3}
                              \and Martina Kleinheinrich \inst{1}
                              \and Zoltan Kovacs \inst{1}
                              \and Lutz Wisotzki \inst{4}}
                              
\offprints{K. Meisenheimer, \email{meisenheimer@mpia.de}}

\institute{Max-Planck-Institut f\"ur Astronomie, K\"onigstuhl 17, D--69117 Heidelberg, Germany
           \and Department of Physics, University of Oxford, Denys Wilkinson 
           Building, Keble Road, Oxford OX1 3RH, U.K.
           \and School of Physics and Astronomy, Cardiff University, 5 The 
           Parade, Cardiff CF24 3YB, U.K.
           \and Astrophysikalisches Institut Potsdam, An der Sternwarte 16, 
           D--14482 Potsdam, Germany
           \and Astronomisches Rechen-Institut, M\"onchhofstr. 12-14, D--69120 
           Heidelberg, Germany}

\authorrunning{Borch et al.}
\titlerunning{The stellar masses of 25000 galaxies at $0.2 \le z \le 1.0$}

\date{Received 19 October 2005 / Accepted 25 March 2006}

\abstract{ We present an analysis of stellar 
mass estimates for a sample of 25000 galaxies from the COMBO-17
survey over the interval $0.2 < z < 1.0$. We have developed, 
implemented, and tested 
a new method of estimating stellar mass-to-light ratios, 
which relies on redshift and spectral
energy distribution (SED)
classification from 5 broadband and 12 medium band filters.  
We find that the majority ($>60\%$) of massive galaxies 
with $M_* > 10^{11} M_{\odot}$ at all $z < 1$ 
are non-star-forming; blue star-forming galaxies dominate
at lower masses.  We have used these mass estimates to explore the 
evolution of the stellar mass function since $z =1$.  
We find that the total stellar mass density
of the universe has roughly doubled since $z \sim 1$.
Our measurements are consistent with other measurements
of the growth of stellar mass with cosmic time and 
with estimates of the time evolution of the cosmic star formation rate.  
Intriguingly, the integrated stellar mass of blue
galaxies with young stars has not significantly changed since 
$z \sim 1$, even though these galaxies host the majority
of the star formation: instead, the growth of the 
total stellar mass density is dominated by the growth of 
the total mass in the largely passive galaxies on the red sequence.

\keywordname{Galaxies: mass function -- Galaxies: evolution}
}
\maketitle

\section{Introduction}

Current models for the formation and evolution of cosmic structure
and galaxies provide us with a picture of hierarchical structure
formation, where large objects are built-up successively from the
merging of smaller, previously formed objects (e.g. Blumenthal et
al 1984; Davis et al. 1985) and from the more gradual inflow of gas
(e.g. White and Rees, 1978; Murali et al. 2002). The dynamical
aspects of CDM-based galaxy formation can by now be well described
either by numerical ab-initio simulations (e.g.  Springel \& Hernquist 2003;
Weinberg et al. 2004) or by semi-analytic
modeling (Kauffmann et al. 1993; Cole et al.\ 1994; Cole et al.\ 2000; 
Somerville et al. 2001). Yet, when, how and how efficiently stars formed in the
course of this evolution appears to depend sensitively on
regulating feed-back and on sub-grid (i.e. unresolved) physical processes; consequently, the current generation of models has limited
power to robustly address this most central question of galaxy formation.

In this context, an empirical assessment of the build-up of stellar mass
from early epochs to the present is an important constraint. 
To explore the efficiency of star-formation and feed-back on different mass
scales, one needs to derive not only the integrated stellar mass
density $\langle \rho_*\rangle(z)$ as a function of epoch, but also
the evolution of the stellar {\it mass function} of galaxies. 
At the present day, it has become clear that stellar mass correlates
strongly with a broad range of galaxy properties, such as 
mass density, mean stellar population age, metallicity, 
or alpha enhancement (e.g., Bender, Burstein \& Faber 1993;
Kauffmann et al.\ 2003; Tremonti et al.\ 2004; Thomas et al.\ 
2005).  Study of how some of these scaling relations evolve
with redshift will give still keener insight into 
star formation, feedback, and galaxy assembly through merging.

Over the last decade (e.g., Lilly et al. 1996; Madau et al.
1996) the global build-up of stellar mass has been traced by
estimating the mean star-formation rate (SFR) as a function of
cosmic epoch,
 $\langle {\rm SFR}(z)\rangle$.
These estimates have been based on either the UV luminosity (e.g.
Steidel et al. 1999; Giavalisco et al. 2004, Schiminovich et al.
2005), on the resulting emission line excitation (e.g. Yan et al.\ 1999; 
Hippelein et al. 2003; Brinchmann et al.\ 2004) or on 
thermal infrared or radio emission (e.g., Flores et al.\ 1999;
Haarsma et al.\ 2000; Le Floc'h et al.\ 2005).
The UV- and line emission-derived 
estimates require large and uncertain dust extinction 
corrections (e.g., Meurer et al.1999; Adelberger and Steidel 2000); 
thermal infrared and radio emission are less dust-sensitive but
are currently limited in sensitivity to probing 
intensely star-forming galaxies at intermediate
and high redshift.  The latter approach
therefore requires large corrections for the faint end of the
IR or radio luminosity function to estimate $\langle {\rm
SFR}(z)\rangle$. Both approaches rely quite sensitively on a prior
assumption for the stellar initial mass function, as the
bolometric luminosity is dominated by the UV radiation from $\sim 5$M$_{\odot}$
stars, while the mass budget is made up by stars of solar
mass and below. Finally, even perfect knowledge of the star formation
density function at different redshifts, $\rho({\rm
SFR},z)$, does not contain enough information to predict the
observable stellar mass function of galaxies, p$({\mathrm
M}_*,z)$.

A conceptually obvious, but practically quite difficult, path to
bypass these problems in estimating the build-up of stellar mass,
is to estimate p$({\mathrm M}_*,z)$ directly from stellar mass
measurements at different redshifts. Two observational avenues
present themselves, dynamical mass estimates from kinematic
measurements at different redshifts, and estimates of the stellar
mass-to-light ratio ($M/L$) based on the spectral energy
distribution (SED) of a stellar population. Dynamical mass
estimates from molecular gas kinematics have now been demonstrated
to redshifts $z>2$ (e.g. Neri et al.\ 2003), but will remain
restricted in the immediate future to exploring the relatively
small fraction of massive, gas-rich galaxies. Kinematics of
galaxies to $z\ge 1$ are now routinely being measured, both from
absorption lines (e.g., Holden et al.\ 2005; 
van der Wel et al.\ 2005) and emission line
rotation curves (Vogt et al.\ 1996, 1997; B\"ohm et al.\ 2004; 
Conselice et al.\ 2005); current published sample sizes
are in the 10s to 100s.

Stellar luminosity is the simplest observational proxy for 
stellar mass, but cannot offer stellar masses with 
relevant precisions (better than about a factor of two),
as the rest-frame
optical $M/L$ values of stellar populations with different SFHs
can vary by more than an order of magnitude.  Rest-frame 
near-infrared $M/L$ ratios vary less as a function of 
star formation history; nevertheless, different plausible star formation histories differ still substantially in their rest-frame infrared
$M/L$ (by a factor of four
or more).

However, the luminous properties of galaxies can be used
to estimate stellar masses.  Low-resolution (R $\sim 3-30$) 
galaxy spectral energy distributions (SEDs) 
do not contain enough information to robustly infer
the  SFH, metallicity, or dust content of
a galaxy.  Older stellar ages, higher metallicity, or
more dust extinction all make SEDs redder.  Yet,
all these effects not only
redden the SED, but also lower the emergent optical flux (at a
given mass) by similar amounts. Hence, stellar $M/L$ can be estimated from
SED-template fits, the strength of the 4000\AA - break (D4000) or
even the (B-V)$_{\rm rest}$ color, for a wide range of the above
parameters (e.g., Brinchmann \& Ellis 2000; Bell \& de Jong 2001;
Papovich et al.\ 2001; 
Kauffmann et al.\ 2003). If spectra are available, 
this approach can be refined
to include two or more
spectral diagnostics, e.g. D4000 and EW(H$\delta$) (Kauffmann et al
2003); indeed, recent works make use of  
higher-resolution spectra to robustly estimate
stellar mass (e.g., Panter et al.\ 2004). 
Stellar masses estimated in these ways are consistent
at the $\sim 0.1$ dex level (e.g., Bell et al.\ 2003;
Drory et al.\ 2004b; Salim et al.\ 2005) and correlate
strongly with dynamically-measured masses (e.g., Bell \& de Jong 2001;
Drory et al.\ 2004a; van der Wel et al.\ 2005; Cappellari et al.\ 2005).
One has to be aware, however, that very dusty galaxies or galaxies with  SFHs featuring by large and recent (less than 2\,Gyr ago) bursts of star formation will have poorly-measured masses 
($\ga 0.3$\,dex error) if not derived from template spectra accounting explicitly for these effects.

SED or spectra-based $M/L$ estimates have been used to describe the
local galaxy population (Cole et al.\ 2001; 
Bell et al.\ 2003; Kauffmann et al.\ 2003),
based on samples of many thousands of galaxies, and on modest samples of
distant galaxies $z\ge 1$, where deep near-IR data have permitted
measurement of the optical rest-frame SEDs (e.g. Papovich et al.\ 2001,
Rudnick et al.\ 2003, Drory et al.\ 2004a, 2005; Fontana et al.\ 2004).  
Studies of distant $z \ge 0.5$ samples have been 
hampered by a number of largely inevitable difficulties: 
small number statistics for the most massive galaxies, 
field-to-field variations (as quantified elegantly by Drory et al.\ 2004a
and Somerville et al.\ 2004), and increased stellar mass uncertainties
due to limited spectral coverage and/or the use of photometric redshifts.

The purpose of the present paper is to derive SED-based stellar
mass estimates for all galaxies in the COMBO-17 sample (Wolf et al.\
2001,2003).   We quantify field-to-field, number statistics, 
and photometric redshift uncertainties as carefully as possible.
We then use those mass estimates for a
comprehensive estimate of the global stellar mass density in the
redshift range $1>z>0.2$, as well as of the galaxy mass function
over the same period.  
This paper is organized as follows. Section \ref{data} briefly
recapitulates the data base of the COMBO-17 survey and outlines its redshift and SED determination. Section \ref{masses} describes the methodology 
for estimating galaxy masses and discusses the errors involved. Section \ref{results}  presents our results on the evolution of the stellar mass in galaxies 
since $z = 1$.  In Section \ref{discussion} we discuss these results in
respect to previous work, the cosmic star formation history  and
models of galaxy formation.  We assume $H_{0} = 70\,h_{70} $\,km\,s$^{-1}$\,Mpc$^{-1}$; 
$\Omega_\Lambda=0.7$ and $\Omega_{\rm m}=0.3$ and adopt $h_{70} \equiv 1$.

\section{The galaxy sample}
\label{data} 
Our analysis is based on the optical multi-color images in three fields (CDFS, Abell 901, 11h field) of the COMBO-17 survey (\cite{Wolf_2003}).  We use
the photometric catalog of Wolf et al. (2003), in the revised version presented by Wolf et al. (2004) to which we refer for further details. Each square field covers $\sim 950\,\sq\arcmin$ on the sky, corresponding to a co-moving
transverse size of $\sim 22$ Mpc at z$\sim 0.7$. The three fields
are so widely separated as to be completely uncorrelated. The
combination of deep observations in 5 broad bands (UBVRI) and 12
medium band filters between 400 and 930 nm, allows us to determine
very accurate photometric redshifts (see Section\,\ref{multicolor},
below and Wolf et al.\ 2004)  
and accurate UV--optical spectral energy distributions of
galaxies (Section\,\ref{SED_lib}). Together, the three fields
contain more than 26\,000 galaxies with redshift
estimates out to $z = 1.07$ which is the limit of reliable, precise
redshift estimates based on optical filters only.

The multi-color photometry can be used to {\it a)} classify each
object as star, galaxy, QSO or indeterminate,
  {\it b)} estimate its redshift, e.g. if it is a galaxy, and {\it c)} fit a
template SED to
the fluxpoints, and derive physical parameters, such as the
stellar $M/L$.  In principle, such a process can be carried
out in one step (e.g., Rudnick et al.\ 2003) where classification, redshift
and stellar mass are estimated jointly.  However, experimentation 
with COMBO-17 data showed that physically-motivated
template sequences with composite (old+young) stellar populations 
gave poorer redshift accuracy than the Wolf et al. (2004) 
dust-reddened single-burst template set.
Consequently, in this paper we have adopted a two step classification process.
First, we classify the objects and estimate redshifts 
using dust-reddened single burst templates 
(Wolf et al.\ 2004), which have been proven to give
the best photometric redshifts as tested against a large spectroscopic 
training dataset, but clearly cannot provide the most realistic 
stellar masses.  
Therefore, the stellar masses of all galaxies and their uncertainties
are obtained by fitting the optical 17-band SEDs with a new set of 
templates with more plausible SFHs.

\subsection{Object classification and redshift estimates}
\label{multicolor} 
In  COMBO-17, objects are classified by their
location in the 16-dimensional color space  (\cite{Wolf_2001}).
We use four broad-band color indices:  $U-B$, $B-V$, $V-R$, 
$R-I$, and 12 medium band indices which always relate to the closest
broad band filter.
Color libraries of four object classes, namely main sequence
stars, white dwarfs, galaxies and quasars, are used as templates.
The galaxy templates span the redshift range $0 < z < 1.4$, the
quasars the range $0 < z < 5$. In the first step, each object is
assigned to one of these four types by comparing its observed
colors (and their errors) to the color distribution of the
templates. For an unambiguous  classification, we require a
relative probability of $p_i/\sum{p_i} \ge 0.75$. In fact, $
98\%$ of all objects at R $< 24$ are unequivocally assigned to one class by
this requirement. The object classification relies primarily on colors;
morphological information is used only for objects with $p_i/\sum{p_i} < 0.75$ for all classes.

For galaxies and quasars,  we derive the joint probability of a
given redshift and a given rest-frame spectral energy
distribution (SED), accounting for photometric errors. The galaxy
SEDs are represented by a set of simple templates (single-burst
stellar populations with reddening) that was tuned to provide the
most robust redshift estimates through comparison with
spectroscopic redshifts (see \cite{Wolf_2004}). This procedure
provides a {\it minimum error variance} estimation of both the redshift
and the SED template. 
For this paper, galaxies with well-measured minimum-error-variance 
redshifts are used, with an additional R-band aperture
magnitude cut R$_{\rm ap} < 24$ to ensure that only galaxies
with reasonably reliable redshifts are included in the analysis.

Wolf et al. (2004) have shown that in the redshift range $0.1 < z < 1.05$ the redshift accuracy of COMBO-17 is well approximated by
$$ \sigma_z = 0.007 (1+z) \sqrt{1 + 10^{0.8(R-21.6)} } , $$
where $R$ is the central R-band magnitude of the galaxy (measured
in a 1\farcs 5 aperture), and does only weakly depend on galaxy color (SED).

\subsection{Determining the SED of the galaxies}
\label{SED_lib}

Obviously, the single burst templates used for the redshift
estimates hardly can account for the wide variety of SFHs in galaxies. Detailed
observations of resolved stellar populations
(e.g., Rocha-Pinto et al.\ 2000; Harris \& Zaritsky 2004) 
and integrated spectra (e.g., Trager et al.\ 2000; 
Proctor \& Sansom 2002; Panter et al.\ 2004) show that most
galaxies have formed stars over extended periods --- oftentimes
over almost a Hubble time.  
As old stars usually dominate the mass while young
stars often dominate the optical flux, it is essential for
estimating the mass-to-light ratio $M/L$ to constrain their
relative amount as accurately as possible from the observed SED.

\begin{figure}
  \resizebox{\hsize}{!}{\includegraphics{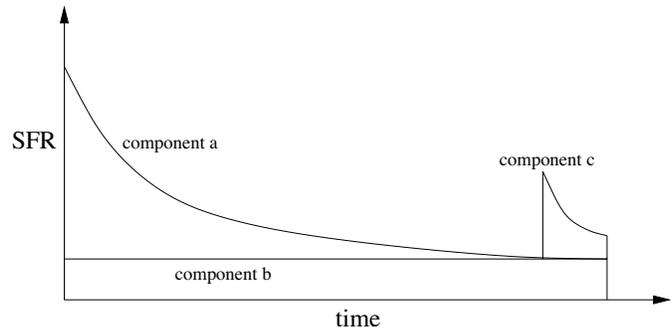}}
  \caption{Parameterization of the star formation history in our templates.
An exponentially decaying star formation rate simulates an initial
burst (component a). A second component b provides some constant star formation 
since the initial burst. For starburst galaxies
a second burst in the recent past is assumed (component c).   }
\label{SED_model}
\end{figure}

Adopting the classification and redshifts from the simple
templates, we estimate \mlstar\ with a second SED
template set that reflects more plausible SFHs. Specifically, we have devised a  library of 100 spectral
energy distributions built with the PEGASE-code (\cite{pegase}), 
sorted from red to blue by an SED index $S$, $0 < S <
99$, which  reproduces the sequence of mean UV-optical
spectra of nearby galaxies collected by \cite{Kinney}.\footnote{In
this way, the new library closely matches the template library
which has been used in early implementations of the multi-color
classification (see Wolf et al.\ 2001; 2003).}
The underlying SFHs have been parameterized by the three-component model
depicted in Figure \ref{SED_model}. For the "old" population, we
assume an exponentially decaying SFH (with 
$e$-folding time 1 Gyr)
several billion years ago
(component $a$),  plus a constant star formation rate since
the initial burst (component $b$). To reproduce the empirical SEDs between E galaxies and
Sb galaxies in \cite{Kinney} we continuously decrease the age
of initial burst by a factor of 2.5 and increase the $b$ component to
a maximum of $b/a = 0.16$\footnote{Parameters $a$, $b$, and $c$ 
are defined in terms of total stellar mass contained in each component; 
see, e.g., Equations 4 and 5. }. 
For even bluer galaxies (the "starburst
galaxies" SB6 to SB1 in \cite{Kinney}) we include a second burst in the
recent past (60 million years ago, component $c$). Its
exponential decay time is kept constant, $\tau_c = 10^8$ yrs.
Bluer galaxies are generated by increasing the relative amount of
recent star formation $c/a$ according to the prescription 
$$ c/a = 0.002 ~e^{S / 17.2} - 0.021, $$
in the range $40 \le S \le 99$.

For component (a) we chose to follow the chemical evolution 
of the model galaxy using a closed box model, starting at 1/20
solar metallicity, as implemented in PEGASE.
The metallicity of components (b) and (c) are 
chosen to be constant at solar metallicity. 
As we assume no infall and no galactic winds
and start at 1/20
solar metallicity, our models will exhibit a significant G-dwarf
problem.  Furthermore, super-solar 
metallicities cannot be easily reached
by this model; thus, the
reddest SEDs (E, S0 type) can only be reproduced using unreasonably
old ages of 20 Gyrs. Therefore, we regard the "ages" as meaningful
 only in a relative sense, with the age of the component a of the
Sb galaxy found to be about $0.4\times t_{max} = 8$\,Gyr (where $t_{max}$
refers to the age of the oldest stars in E/S0 galaxies)\footnote{ 
It is important to note that, owing to the age/metallicity degneracy for 
old stellar populations, the modest changes in stellar M/L caused 
by the old ages of the $S<40$ templates are cancelled out by the 
substantially sub-solar metallicities of these templates.  Thus, the 
template color--stellar M/L ratio correlation is in fact very similar to a 
solar metallicity, 7\,Gyr old population for the $S<40$ templates; such 
ages and metallicities are reasonably appropriate for modelling the 
$0.2<z<1.0$ galaxy population.}.
We have adopted a Kroupa IMF (\cite{Kroupa}) and the
mass regime 0.1-120 $M_\odot$. The stellar $M/L$ values and colors
from such an IMF are very similar to those of 
a Kroupa (2001) IMF or a Chabrier (2003) IMF.  
Dust extinction is crudely accounted
for using PEGASE (see Fioc \& Rocca-Volmerange 1997 for  
details).  In the code the dust optical
depth is estimated from the gas mass and metallicity, scattering
is accounted for crudely following Calzetti et al.\ (1994). For component 
a we choose spheroidal
geometry, while 
a plane-parallel slab geometry, averaged over all inclinations, is chosen for components b and 
c.  This dust prescription reproduces
many of the broad trends observed in local galaxies
(e.g., Wang \& Heckman 1996) reasonably well\footnote{The 
color effects of age, metallicity and dust 
are largely degenerate when estimating stellar $M/L$ from 
broad-band SEDs, therefore the detailed choice of this 
or a different dust model  
does not significantly affect our derived stellar masses.}.

\begin{figure}
  \resizebox{\hsize}{!}{\includegraphics{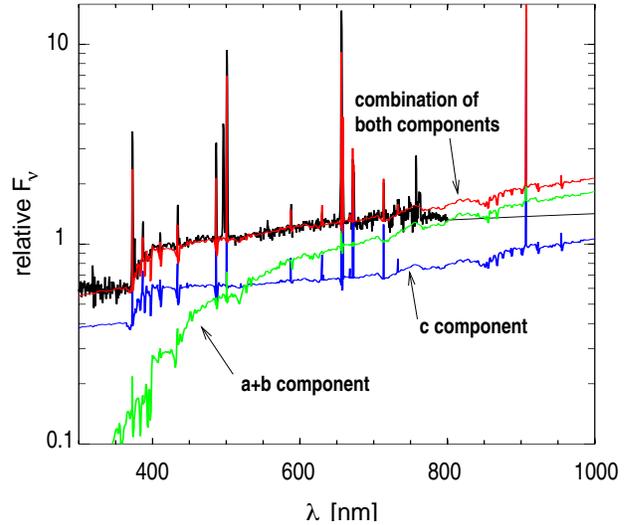}}
  \caption{An example for the template spectra for a starburst galaxy.
The original Kinney et al. spectrum (SB-2) is shown and our best
fit from the PEGASE spectra. It is composed of an old stellar
population ($a + b$ component in Fig.\,\,\ref{SED_model}) and a
young stellar population ($c$ component).
  }
\label{SB-beispiel}
\end{figure}

\begin{figure}
\resizebox{\hsize}{!}{\includegraphics{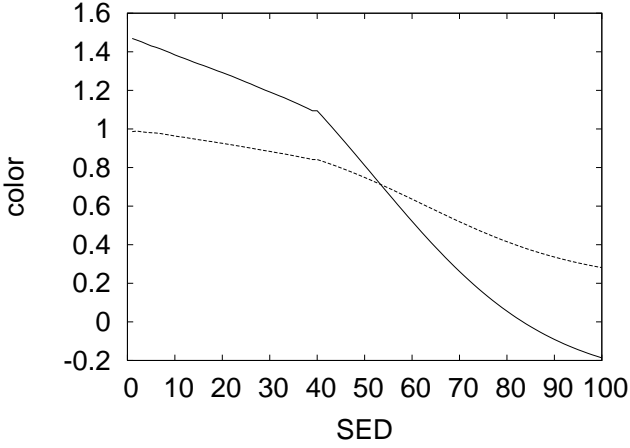}}
  \caption{Rest-frame colors U-V (solid line) and B-V (dotted line)
as function of the SED type $S$
in our library of galaxy templates.} \label{SED_color}
\end{figure}

In order to be suitable for multi-color classification, the 
one-dimensional library $S(0\dots99) = S(a,b,c)$ has to be
monotonic both in color space and in the space of its astrophysical
parameters: $b/a$, $c/a$ and the age of the old stellar
population. Figure \ref{SB-beispiel} shows an example of one
starburst galaxy from \cite{Kinney} and its composition from an
old plus a young stellar population. For comparison with previous
work we characterize the SED types $S$ also by their rest-frame
colors U-V and B-V (see Fig.\,\ref{SED_color}).

The best-fitting SED is
obtained by comparing the observed colors of each galaxy with the
colors of the SED library spectra, shifted to the most likely
redshift $z$ as measured using the Wolf et al.\ (2004) classifier 
(Section \ref{multicolor}).

\section{Estimating stellar masses}
\label{masses}

Based on the $(M/L)_{*}$ of the best-fitting new SEDs, stellar
masses can now be estimated. We choose to normalize the stellar
mass by the reddest well-measured filter to minimize the 
influence of bursts of star formation.  The
medium-band filter which provides the best compromise between
longest observed wavelength and signal-to-noise ratio is $816/21$,
with a central wavelength of 816 nm and a width of 21 nm. The
$816/21$ filter has a limiting magnitude of 22.9 ($10\sigma$) in all
of our fields. The lower edge of the $816/21$ filter, $\lambda_{\rm min}
= 804$\,nm, sets the maximum redshift where the stellar mass can be
measured reliably: for a significant light contribution from the
old stellar population, one needs to make sure that only flux at
rest-frame wavelengths $\lambda_{\rm rest} > 400$\,nm is sampled.
Hence, we restrict our current investigation to redshifts $z \le 1.0$.

In practice, we convert the observed quantities $z,~S, ~f_{816}$ of each individual
galaxy into stellar mass in the following way:

\noindent
(1) 
The luminosity delivered from the
PEGASE code, which is normalized to 1 $M_\odot$, is used to calculate
the apparent flux $f_{\lambda}$. For simplicity, we first consider 
only one component (luminosity $F_{\lambda}$). 
In this case the observed flux of a galaxy
of stellar mass $M$ is given by:

\begin{equation}
\label{pegase}
f_{\lambda} =
   \frac{\int F_{\lambda} \cdot 
              Q_{\lambda} \, d\lambda}
        {\int Q_{\lambda} \, d\lambda}
    \cdot \frac{\frac{1}{m} \frac{M}{M_\odot}}{4 \pi D_L^2(z) (1+z) /h_{70}^2} .
    \end{equation}
Here $m \cdot M_\odot$ denotes the {\it stellar} mass of the PEGASE template. The
integral runs from 0 to $\infty$. $Q_\lambda =  q(\lambda) t(\lambda) $ is the
spectral response of the system, in our case given by the transmission of the 816/21
filter and the quantum efficiency of the CCD.
The factor $(1+z)$ accounts for the different wavelength scales of the observed flux
$f_{\lambda}$ and the rest-frame luminosity $F_{\lambda}$.
Both $F_{\lambda}$ and $f_{\lambda}$ are given in 
${\rm W\, m^{-2} \, nm^{-1}}$. 

In practice, COMBO-17 measures the central flux $f_{\gamma,ap}$ of each object in 
photon units $\gamma {\rm m^{-2}s^{-1} nm^{-1}}$ (\cite{Wolf_PhD}) after convolving all
images to a common effective PSF of 1.5 arcseconds. Thus, for well resolved galaxies, the
central flux $f_{\gamma,ap}$ has to be corrected to the total flux $f_{\gamma,tot}$ by a correction factor $\eta_{tot}$. It is derived for each galaxy from the SExtractor {\tt MAG-BEST} magnitudes (Kron magnitudes) on our very deep R-band image: $\eta_{tot} \equiv 10^{0.4 (R_{ap} - R_{\rm Kron})} $. Since we ignore wavelength dependent aperture effects, the apparent flux $f_{\lambda}$ in equ. (\ref{pegase}) is approximated by:

\begin{equation}
\label{umrechnung}
  f_{\lambda} = \frac{hc}{\lambda} f_{\gamma,tot} = \frac{hc}{\lambda} f_{\gamma,ap}
\cdot \eta_{tot} 
\end{equation}
With equation (\ref{umrechnung}), setting $f_{\gamma,ap} = f_{816}$, and solving
formula (\ref{pegase}) for the mass one gets:
\begin{equation}
\label{zwischenergebnis}
 \frac{1}{m} \frac{M}{M_\odot} = 
 \frac{f_{816} \, \eta_{tot} \cdot 4 \pi D_L^2(z) \cdot (1+z) \,\frac{hc}{\lambda}
 \int Q_{\lambda} \, d\lambda}
{h_{70}^2 \cdot \int F_{\lambda} \cdot Q_{\lambda} \, d\lambda}
\end{equation}

\noindent (2)
As our templates are composed of three components $a,b,c$, both the
luminosity and the stellar mass have to be synthesized:
For each SED type $S_i = S(a,b,c)$ the total luminosity (per $M_\odot$) is
\begin{equation}
\label{spektrum_linaerkomb}
F_{\lambda} = a \cdot F_{\lambda,a} + b \cdot F_{\lambda,b} +  c \cdot F_{\lambda,c}.
\end{equation}
Similarly, the total stellar mass results from a linear combination of the
individual components with the parameters $a$, $b$ and $c$:
\begin{equation}
\label{masse_abkuerzung}
  m = a \cdot m_a + b \cdot m_b + c \cdot m_c.
\end{equation}
Here $m_a$, $m_b$ and $m_c$ are the {\it stellar} mass fractions of the respective components.
Inserting (\ref{masse_abkuerzung}) and (\ref{spektrum_linaerkomb}) into
equation (\ref{zwischenergebnis}), yields the final result:
\begin{equation}
\label{final}
 h_{70}^{2} \frac{M}{M_\odot} = \frac{ 
f_{816} \, \eta_{tot} \cdot 4 \pi D_L^2(z) \cdot (1+z) \,
   \frac{hc}{\lambda} \int Q_{\lambda} \, d\lambda}
{\sum_{k=a,b,c}{\frac{k}{m_k} \int F_{\lambda,k} \cdot Q_{\lambda} \, d\lambda} }.
\end{equation}
Equation (\ref{final}) delivers the stellar mass of any galaxy with
observed values $f_{816}, \, \eta_{tot}$, for which the classification has succeeded in determining the redshift $z$ and the spectral type $S = S(a,b,c)$. The $< 2\%$ of galaxies for which no reliable redshift can be determined are faint and blue: mis-placing them does only affect the mass function of blue galaxies below $3\,10^9$\,M$_{\sun}$ (see Fig.\,\ref{massfctbr}). 

\subsection{Intrinsic errors}
\label{fehlerrechnung}
As obvious from equation (\,\ref{final}), the intrinsic errors of the stellar masses depend both on the errors in 
redshift $z$ ({\it  via} $(1+z)D_L^2(z)$) and in SED type $S$ ({\it i.e.} the $M/L$ ratio: $M/F_\lambda$). Since a red SED at lower redshift can produce similar colors as a blue SED at slightly higher redshift, the errors in $z$ and $S$ are correlated. 
To account properly for this dependence we calculate 
the covariance matrix for each galaxy. To this end, besides determining the SED type $S(z_0)$ for the optimum redshift (according section
\,\ref{multicolor}, with its error $\sigma_z$), we let $z$ vary in the interval
$[z_0-2\sigma_z,z_0+2\sigma_z]$. From the corresponding points $S_i(z_i)$. 
we calculate the covariance matrix 

$$C = \left( 
\begin{array}{cc}
P_i^2 (\overline{z_i - z_0} )^2 & P_i^2 
(\overline{S_i - S(z_0})( \overline{z_i - z_0}) \\
 & \\ P_i^2
 (\overline{z_i - z_0})( \overline{S_i - S(z_0)}) &
P_i^2 (\overline{S_i - S(z_0)})^2 
\end{array} \right),
$$
where each point is weighted by the
probability 
\begin{equation}
P_i = exp \left[ -\frac{1}{2} \left( \frac{z_i - z_0}{\sigma_z} \right)^2
\right].
\end{equation}
This matrix has to be diagonalized to obtain independent
new parameters $\zeta$ and $\mu$, calculated for each galaxy from the
parameters $z$ and $S$. The mass error is then estimated by 
gaussian error propagation from these new parameters and the error of the
observed flux, $\sigma_f$:
\begin{equation}
  \sigma_M = \sqrt{
      \left( \frac{\partial M}{\partial \zeta}\sigma_\zeta \right)^2
    + \left( \frac{\partial M}{\partial \mu}  \sigma_\mu \right)^2
    + \left( \frac{\partial M}{\partial f_{\gamma,ap}}\sigma_f \right)^2} .
\end{equation}

\subsection{Calibration of mass scale and systematic errors}
\label{systematics}

\begin{figure}
  {\includegraphics[width=92mm]{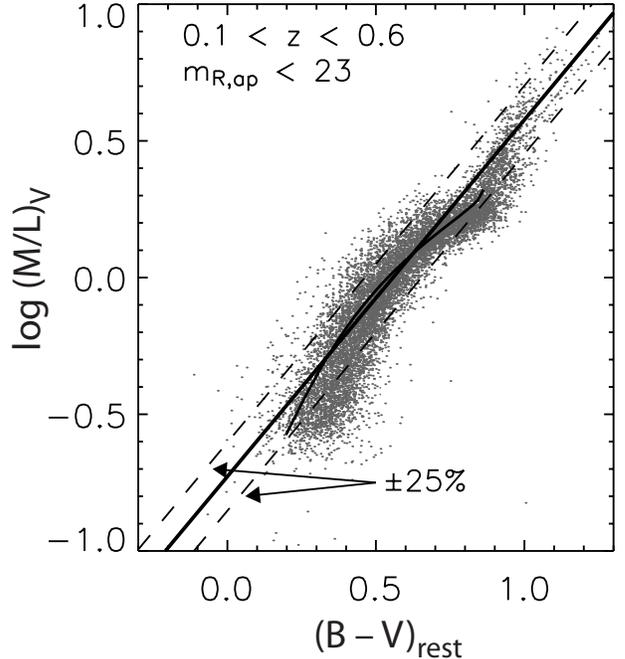}}
  \caption{V-Band mass to light ratio {\it versus } (B-V)$_{\rm rest}$ color of
COMBO-17 galaxies. For comparison the Bell et al.\ (2003) color--$M/L$
correlation is 
shown as straight line. The transition between early type
($S < 40$) and blue galaxies occurs at $(B-V)_{\rm rest} = 0.8$.
Also overplotted is the color track of a sequence of exponential star formation 
histories with solar metallicity and age 7\,Gyr (solid curve).  }
  \label{ML_BR}
\end{figure}

To investigate the global (average) properties of hundreds or
thousands of galaxies within a redshift range, the statistical
mass error of an individual galaxy is less important than any
systematic error in the mass scale. We identify three main sources
of systematic errors:
\begin{enumerate}
\item Overall calibration of the mass-to-light ratio scale. 
\item SED dependent errors caused by an incorrect conversion from observed colors of a galaxy into an SED $S$ (i.e. essentially: assuming an incorrect SFH which likewise fits the colors).
\item Redshift dependent errors, which could be either caused by sampling different rest-frame wavelengths or by using a common set of template spectra
for a galaxy population which evolves with cosmic time.
\end{enumerate}
In order to estimate how much these systematic 
effects might influence our results, we {\it first}  
compare (Fig.\,\ref{ML_BR}) the color--$M/L$ relation from 
Bell et al.\ (2003) which is  an update of Bell \& de Jong (2001), 
scaled to a Kroupa et al.\ (1993) IMF, 
with the result of our mass
estimation for all galaxies at $0.1 < z < 0.6$ (where the
rest-frame V is sampled by the COMBO-17 photometry), and with
$R$-band aperture magnitudes R$_{\rm ap} < 23$ (to avoid excessive statistical errors). The discrepancies both at the blue and red end can be easily
understood by the fact, that 
our template age-metallicity combination 
(red: metal poor and unreasonably old; blue: solar metallicity and 
young $\la 0.1$\,Gyr) should result in stellar $M/L$ ratios
at a given optical color which are very slightly smaller than those
of Bell \& de Jong (2001), who assumed 12\,Gyr old populations
and relatively high metallicities.  Notwithstanding this 
slight difference, we conclude
that the stellar $M/L$ ratios used in this paper are very similar
to those used for local studies, arguing that comparisons
of this work to $z \sim 0$ studies should be reasonably robust.
\vspace{.2in}

\begin{figure}
{\includegraphics[width=92mm]{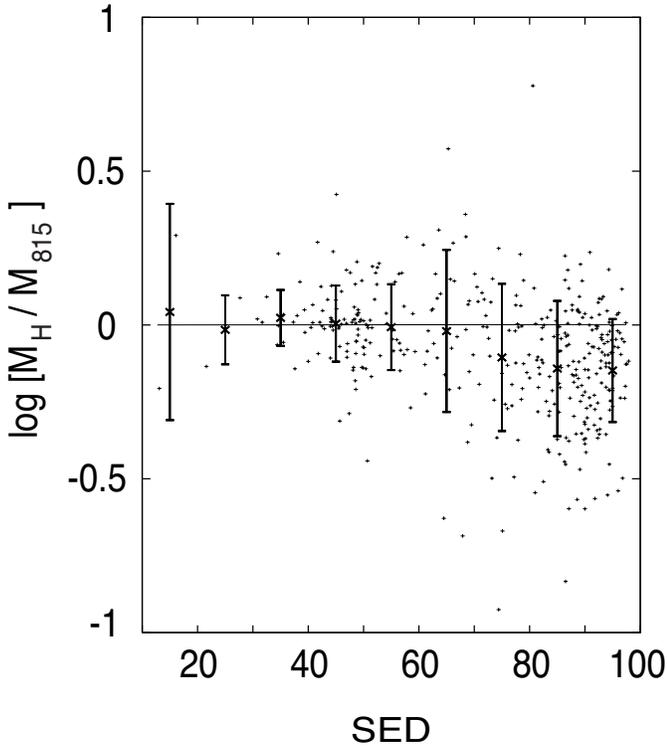}}
\caption{Ratio between the stellar mass estimation based on the
$H$-band and the stellar mass from the 816/21 filter as a function
of SED type. The crosses with error bars show averages in SED bins
($\Delta S = 10$) and their {\it  rms} variation.}
 \label{Hvs815}
\end{figure}


Since our stellar masses are derived from observed-frame optical
data alone,
an independent {\it second} check of the robustness of the stellar mass
estimates as a function of SED type is to compare the optical-only
stellar masses with stellar masses derived from observed-frame 
near-infrared data. 
We have used deep, public, fully-reduced $J$, $H$, and $K$-band data on a 50 square arcminute sub-region of the CDFS
taken with ISAAC at the VLT as a part of the ESO Imaging Survey EIS
(\cite{EIS-paper}). Their limiting
magnitudes are J$<24.8$, H$<23.4$ and $K_s<22.2$ (5$\sigma$, Vega system).
Although this is only a small fraction of one COMBO-17 field, we find a sample of about 400 galaxies for which
both a reliable COMBO-17 classification and near-infrared
photometry are available. 

We feed the ISAAC images directly in our COMBO-17 pipeline and
re-calibrate the measured flux with respect to main sequence stars
in the field.\footnote{This re-calibration accounts only for the
global zero-point of the EIS data. From the main sequence stars we have some indications that the zero-points of the 8 subfields are slightly inconsistent
with each other. }
In order to allow matched photometry, the field
distortions of the ISAAC camera have been corrected radial
symmetrically, based on relative positions of bright objects on
our deep R-band image and the individual J, H, K pointings.

For the comparison, we do not re-calculate best-fit redshifts
and SED types using the optical and near-infrared data.  Instead, 
we use the optically-derived redshifts and SED fits to blindly predict
stellar $M/L$ values for the observed-frame $J$, $H$, and $K$-bands, which 
are then multiplied by the $J$, $H$, and $K$-band photometry to 
derive new stellar masses\footnote{Recall that the default
mass estimates were derived from SED-derived $M/L$s in the 
observed-frame 816/21 passband, multiplied by the 816/21 observed-frame
luminosity.}.  That is, we do not attempt to find a better overall solution for fitting the template spectra to {\it all} observed
colors (16 optical plus 3 including the near-infrared).
This approach has the advantage that we are sensitive only to
the effect due to sampling different parts of the galaxy spectra for estimating
the stellar mass. As the relative fraction of light from young and old stellar
populations is a function of the rest-frame wavelength which is sampled, any
inconsistencies between our synthesized spectra (and their composition) and the true spectra of the galaxies should become evident.

In Figure \ref{Hvs815} the comparison between the mass
estimations from filter 816/21 and from the $H$-band is shown as a
function of SED type. The agreement is on average much better than
0.1~dex (25\%) for early type and spiral galaxies ($S < 70$).
Later templates --- those with significant bursts of star 
formation in our template set with $S > 70$ --- 
are offset by up to 0.15 dex in the sense that
the optically-derived masses {\it overestimate} the stellar mass.
The object-by-object scatter is largely accounted for by random errors
in stellar mass: more than 60\% of galaxies have mass
differences of less than the combined mass uncertainties.
The use of $J$-band for the mass derivation leads to similar results.
If $K$-band is used for the mass derivation, again galaxies at 
$S < 70$ show consistent masses to within their combined uncertainties, 
whereas galaxies with the bluest stellar populations $S > 80$ show
$K$-band derived stellar masses $-$0.3 dex lower than the optically-derived
masses.  While this offset is significant, it is worth bearing in mind
that the masses of the bluest, most strongly star-forming galaxies
were expected to be systematically 
uncertain at this level (owing to the unknown 
frequency of bursts of star formation; Bell \& de Jong 2001; 
Kauffmann et al.\ 2003).  Furthermore, galaxies with $S < 70$ --- whose
stellar masses appear to be consistently derived using either optical
or near-infrared data ---
form more than 80\% of the total integrated stellar mass at all
$z < 1$.  Thus, we conclude that 
the global results presented in this paper would not significantly change
if near-infrared data were used in addition to the optical data to 
derive the stellar mass function evolution. 

\begin{figure}
\centering \resizebox{\hsize}{!}{\includegraphics{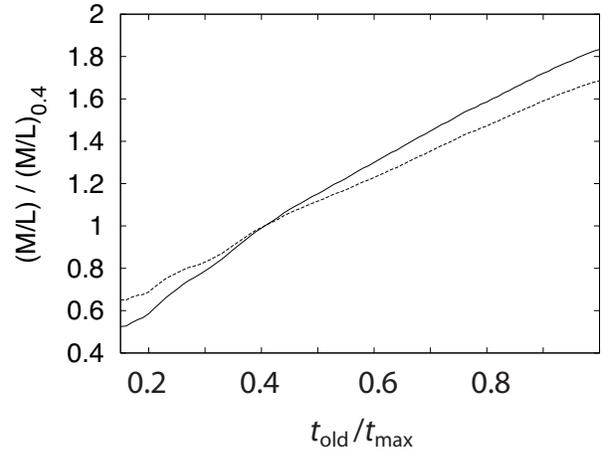}}
\caption{Inferred stellar mass-to-light ratio for starburst
galaxies as a function of the age of the underlying old stellar
population (continuous line: V-band, dashed: K-band $M/L$). The plot
show the ratio between the true $M/L$ ratio and that inferred with
our $t_{old}/t_{max} = 0.4$ templates when classifying starburst
galaxies for which we varied the age of the old stellar
population. Note that both the V-band and K-band $M/L$ ratio would
be affected to a similar extent.  The input template types were
$50<S<90$.
 }
\label{ML_told}
\end{figure}

Although Bell \& de Jong (2001) have demonstrated that there is an
average correlation of the stellar $M/L$ ratio and optical color
for a wide variety of SFHs (see Fig.\,
\ref{ML_BR}), different SFHs
resulting in the same optical color might well differ by a factor
of 2 in their $M/L$ ratio. Our SED library has been constructed
under the assumption that all blue galaxies ($S \ge 40$) possess
a common old stellar population of age $t_{\rm old} = 0.4\, t_{\rm max}$. 
Since at $z = 1$ the universe was half its present age, this
assumption might introduce a redshift dependent systematic error
in the mass of blue galaxies. As a {\it third} check we have investigated this effect by
constructing blue galaxies with various values of $t_{\rm old}$, subsequently classified with our template library (constant $t_{old} =
0.4 t_{\rm max}$).  As one can see from Fig.\,\ref{ML_told}, increasing
the age of the old stellar population by a factor of 2 ({\it i.e.}
from about 5 to 10 Gyrs) would increase the mass in blue galaxies
by about 50\%. On the other hand, assuming a much younger age,
$t_{\rm old} = 0.2 t_{\rm max}$ could lower the mass in blue galaxies by
40\%.  We conclude that this effect introduces a $\sim 0.1$\,dex
uncertainty in the evolution of stellar mass in blue galaxies in the 
4 Gyr from $z = 0.9$ to $z = 0.3$ (the redshift range probed by  our study), and a $\ll 0.1$\,dex uncertainty in the evolution of total stellar
mass over that period (as blue galaxies do not dominate 
the total stellar mass at any redshift).

\begin{figure}
  \centering
  \includegraphics[width=85mm]{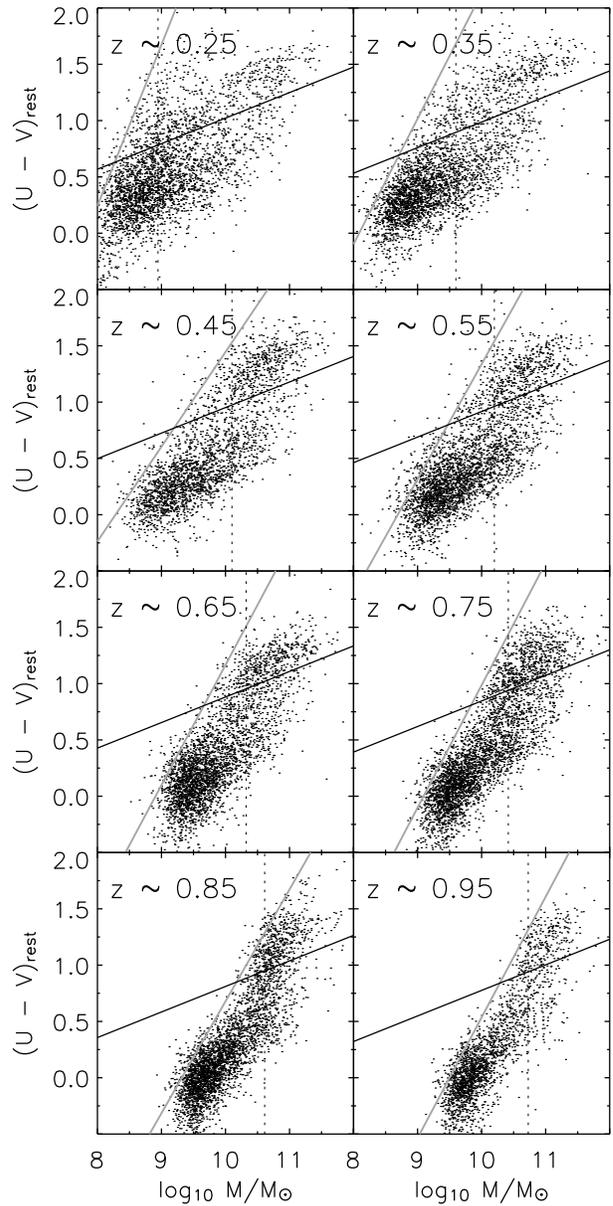}
  \caption{Stellar mass {\it versus} rest-frame U-V color in different
redshift bins. Middle redshift of the $\Delta z = 0.1$ wide bins is given in the upper left of each panel.
The solid black line shows the seperation
between red sequence and blue cloud galaxies (equation (10)).
The grey line indicates the approximate effect of the adopted
apparent magnitude cut ($m_R < 24$) for successful redshift
classification.  The dotted lines indicate completeness limits for red sequence galaxies; blue cloud galaxies are complete well
below this limit.  }
  \label{masse_UV}
\end{figure}

\section{Results}
\label{results}

\subsection{The relationship between color and stellar mass}
\label{mass_UV}

Figure
\ref{masse_UV} shows the stellar mass estimates for the COMBO-17 galaxies
as function of rest-frame U-V color. 
The solid black line indicates the (evolving)
separation of `red sequence' from 'blue cloud' galaxies, proposed by
Bell et al. (2004), after conversion into the \mstar $- (U-V)_{\rm rest}$ plane: 

\begin{equation}
(U - V)_{\rm rest} > 0.227 \log_{10}{\rm M_*} - 1.16 - 0.352 z,
\end{equation}
Not only locally ($z < 0.3$) but also at high
redshift $z \simeq 1$ the intrinsically red (and for the most part
old and non star forming) galaxies constitute over 60\% of the
very massive galaxies (\mstar $> 10^{11} M_{\sun}$).

\subsection{Completeness limits }
\label{mass_complete}

\begin{figure*}
  \centering
  \includegraphics[height=12cm]{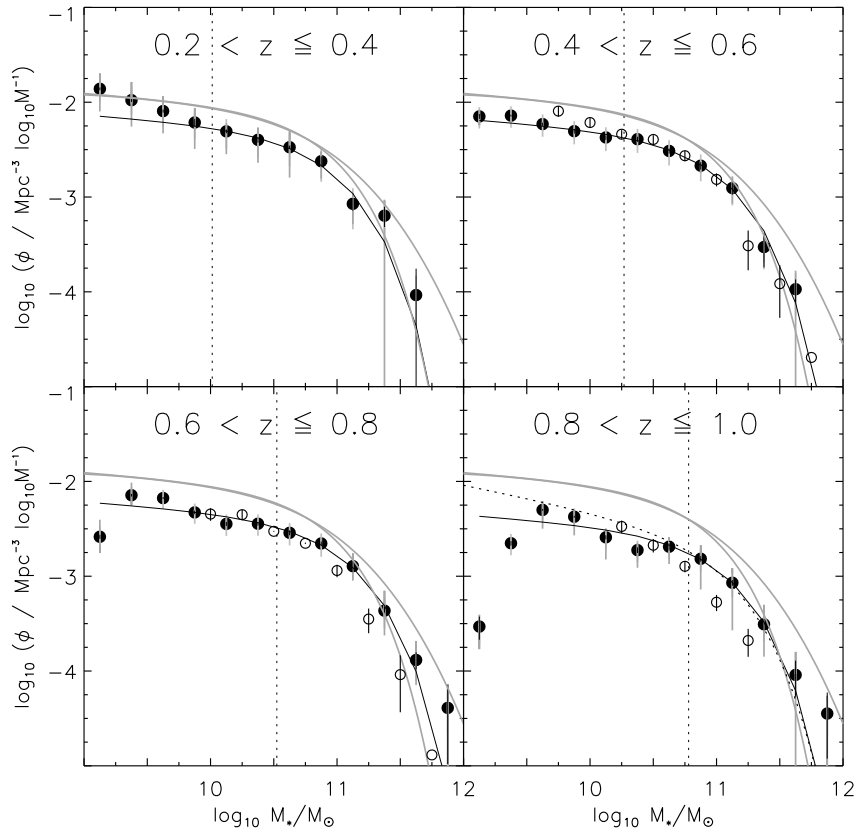}
  \caption{Stellar mass function of all galaxies in different redshift
bins (filled circles). The error bars 
account for Poisson uncertainties
and cosmic variance as estimated from the full range of 
field-to-field variation (dark grey) and 
following Somerville et al.\ (2004; light grey).  
The solid line represents a Schechter function fit (with
constant faint end slope $\alpha$, assumed to be $\alpha = -1.1$ following
the $z \simeq 0$ result of Bell et al.\ 2003). The lower grey line
shows the local mass function; the upper grey curve shows the local 
mass function convolved with a log normal distribution of 
width 0.3 dex to simulate the effect of photometric redshift 
and stellar mass errors on the recovered mass function of high-mass 
galaxies.  The vertical dotted line
shows the completeness limit (for red galaxies) taken from
Fig.\,\,\ref{masse_UV}. Open circles show comparison mass functions
from Drory et al.\ (2004a), and the dotted line in the highest-redshift
bin shows the Schechter fit to the 
stellar mass function from Fontana et al.\ (2004).}
  \label{massfctall}
\end{figure*}

It is obvious from Fig.\,\ref{masse_UV} that the R-band flux
limit of COMBO-17 leads to a mass cut (denoted roughly by the grey line in each panel of 
Fig.\,\ref{masse_UV}), which does not only depend on
redshift but quite strongly on the rest-frame galaxy color:
the least massive red galaxies contained in our sample typically
have $10\times$ higher masses than the least massive blue galaxies.
Thus, the strict mass completeness limit is set by the red sequence
galaxies. We estimate an approximate completeness limit by selecting the mass value above which we find 80\% of our
detected red sequence galaxies in the respective redshift bin
(vertical dotted lines in each redshift panel of Fig.\,\ref{masse_UV}).  
These strict limits are very conservative; in 
particular, the blue galaxy mass function will be complete 
well below this limit.   The same
limits, evaluated for the upper boundary of each redshift bin, are
also shown in the mass function plots (see Figs.\,\ref{massfctall}
\& \ref{massfctbr}).

\subsection{The evolution of the stellar mass function}
\label{mass_function}

We now use the estimated stellar masses of galaxies 
up to $z=1.0$ to derive the evolution of the stellar mass
function.  Since our redshift slices are thin ($\Delta z = 0.2$)
and the limiting mass is evaluated at the upper boundary of each
bin, our samples are effectively volume limited down to 
that limiting mass and we do not use
a $V_{max}$ correction.  We apply individually-derived 
completeness corrections to each detected galaxy 
(see, e.g., Wolf et al.\ 2003 for a full description of 
the completeness properties of COMBO-17).  
These were derived by putting artificial
galaxies into COMBO-17 images and applying the COMBO-17 
detection and classification pipeline.  The completeness
correction was derived as a function of $(z,m_{R,ap},U-V)$ 
by constructing the ratio of number of recovered galaxies 
to by the number of galaxies inserted into the same bin
in redshift, $R$-band aperture magnitude and $U-V$ rest-frame color.
That is, this completeness correction corrects
the number of detected galaxies in a given bin
in redshift, rest-frame colour and magnitude: in practice, this
correction does not exceed 20\% for objects classified as galaxies.   
It does not correct for galaxies not classified by COMBO-17, i.e. which are 
too faint to be classified or have highly unusual 17-passband SEDs.

\begin{figure*}
  \centering
  \includegraphics[height=12cm]{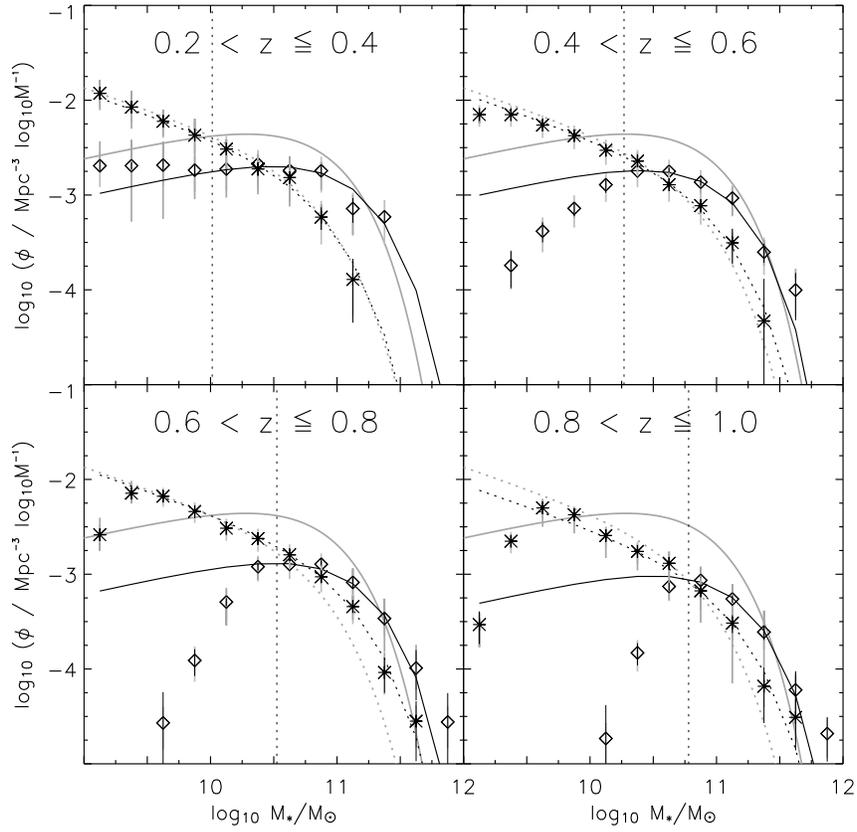}
  \caption{Stellar mass function of {\it red} ($\diamond$) and {\it blue} ($\ast$) galaxies in different redshift bins. The solid black line represents a fit  of the red galaxies with a Schechter function (with 
$\alpha= -0.7$). The Schechter fit to the blue galaxy stellar mass function 
is shown as dotted black curve (assuming $\alpha = -1.45$).  The 
local mass functions, taken from Bell et al.\ (2003), are
shown in grey.  The error bars account for Poisson and 
field-to-field uncertainty.
The vertical dotted line shows the
completeness limit for red galaxies.
}
\label{massfctbr}
\end{figure*}

The results are shown in Fig.\,\ref{massfctall} (total for all galaxies) and Fig.\,\ref{massfctbr} (split into blue and red galaxies). The parameters
of the Schechter fits (shown by the continuous and dotted lines in
Figs. \ref{massfctall} \& \ref{massfctbr}) are given in Table
\ref{Schechter}. Schechter functions are fit to galaxies only
above our conservative completeness limit (as denoted by the dotted line).
The data lack the depth to allow robust estimation of the
faint end slope $\alpha$.  A number of strategies could have been adopted; 
we chose to assume the same values of $\alpha$ as found locally:
$\alpha = -1.1$ (all galaxies), $\alpha = -0.7$ (red galaxies), and $\alpha = -1.45$ (blue galaxies)
following Bell et al.\ (2003)\footnote{Other strategies
could have been adopted for constraining $\alpha$, e.g., adopting
the faint end slopes of the luminosity functions as derived
by Willmer et al.\ (2005).  We chose to adopt values of $\alpha$
fixed to locally-determined values to facilitate comparison 
of mass functions from $z = 1$ to $z = 0$.}.
Mass function parameters for the local
sample are reproduced in Table \ref{Schechter} converted to the cosmology
adopted in this paper and using a universal Kroupa et al.\ (1993) 
IMF (for reference, stellar masses calculated with 
a Chabrier 2003 IMF would be similar
to within a few percent).  

\begin{table}
\caption{Result of the Schechter fits to the mass function. }
\label{Schechter}
\bigskip
\centerline{
\begin{tabular}{l|cccc}
\hline
& & & &\\[-1.5ex]
$z$ & $\phi^* \times 10^4$ & $\log_{10} M^*$ & $\alpha$ & $\log_{10} \rho^*$ \\[.3ex]
\hline
%
\multicolumn{5}{c}{} \\[-1.5ex]
\multicolumn{5}{c}{All galaxies} \\ 
\hline
& & & & \\[-1.5ex]
     0.0 & 35$\pm$4 &  10.91$\pm$0.10  &   $-1.1\pm0.02$ & 8.48$\pm$0.10 \\
     0.3 & 19$\pm$9 &  11.03$\pm$0.08  &   $-$1.1 &   8.34$\pm$0.15  \\
     0.5 & 18$\pm$6 &  11.02$\pm$0.08  &   $-$1.1 &   8.32$\pm$0.11  \\
     0.7 & 16$\pm$5 &  11.09$\pm$0.15  &   $-$1.1 &   8.33$\pm$0.10  \\
     0.9 & 12$\pm$4 &  11.08$\pm$0.11  &   $-$1.1 &   8.17$\pm$0.18  \\
\hline
\multicolumn{5}{c}{} \\[-1.5ex]
\multicolumn{5}{c}{Red sequence} \\ 
\hline
& & & & \\[-1.5ex]
     0.0 & 37$\pm$4 &  10.81$\pm$0.10  &   $-0.7\pm0.07$ & 8.33$\pm$0.10 \\
     0.3 & 17$\pm$7 &  10.97$\pm$0.09  &   $-$0.7 &   8.17$\pm$0.18  \\
     0.5 & 15$\pm$5 &  10.95$\pm$0.10  &   $-$0.7 &   8.08$\pm$0.11  \\
     0.7 & 11$\pm$4 &  11.06$\pm$0.18  &   $-$0.7 &   8.05$\pm$0.10  \\
     0.9 & 9$\pm$3 &  11.01$\pm$0.08  &   $-$0.7 &   7.89$\pm$0.18  \\
\hline
\multicolumn{5}{c}{} \\[-1.5ex]
\multicolumn{5}{c}{Blue cloud} \\ 
\hline
& & & & \\[-1.5ex]
     0.0 & 9.4$\pm$1.0 &  10.80$\pm$0.10  &   $-1.45\pm0.03$ & 7.98$\pm$0.10 \\
     0.3 & 7.5$\pm$3.6 &  10.86$\pm$0.19  &   $-$1.45 &   7.94$\pm$0.15  \\
     0.5 & 7.2$\pm$2.2 &  10.93$\pm$0.12  &   $-$1.45 &   8.00$\pm$0.11  \\
     0.7 & 6.4$\pm$2.2 &  11.07$\pm$0.07  &   $-$1.45 &   8.08$\pm$0.10  \\
     0.9 & 5.3$\pm$1.7 &  11.00$\pm$0.25  &   $-$1.45 &   7.93$\pm$0.10  \\
\hline
\end{tabular}
}
\end{table}

Figure \ref{massfctall} shows clearly that the evolution of the
{\it total} mass function is relatively modest. 
Since the low-mass end is completely
dominated by blue galaxies ({\it  c.f.} Fig.\,\ref{massfctbr}), for
which the completeness reaches to smaller masses, the 
stellar mass function continues to rise well below the formal
mass limit (recall, the mass limit is defined for red galaxies).
We show also near-infrared derived
stellar mass functions from MUNICS (Drory et al.\ 2004a;
open circles) in the $0.4<z\le 0.6$, $0.6<z\le 0.8$, and $0.8<z\le 1.0$ bins,
and the Schechter fit to the K20-derived mass function of 
Fontana et al.\ (2004) in the $0.8<z\le 1.0$ bin.  Overall, the 
agreement between COMBO-17's and other, near-infrared derived $z < 1$ 
stellar mass functions is excellent if one takes into account the large error bars at the bright end (due to small number statistics). In a recent study Bundy et al. (2006) present K-band based mass estimates of 8000 galaxies from the DEEP2 redshift survey (Davis et al. 2003). Although their split between red and blue galaxies, and redshift binning differ from ours, they find very similar results. 

\subsection{The evolution of the stellar mass function of red and blue galaxies}

Owing to the COMBO-17's number statistics and photometric redshift quality,
it is possible not only to constrain the total stellar mass function,
but to also construct mass functions of the 
red sequence and blue cloud separately.  The results
are shown in Fig.\,\ref{massfctbr}.  Diamonds and solid lines
(black: fit to the data, gray: local color-split mass function)
show the stellar mass function of red sequence galaxies.  Asterisks and
dotted lines show the stellar mass function of blue cloud galaxies.
The vertical dotted line shows the position of the completeness
limit for red-sequence galaxies: it is clear from the data points 
that red sequence galaxies
become dramatically incomplete at low masses, whereas the mass function of
blue galaxies continues to rise steeply. 

Focusing on the blue cloud galaxies (asterisks and dotted lines),
it is clear that there is relatively little change in their mass
function since $z \sim 1$.  In this framework, the observed
rapid changes in $L^*$ in the blue galaxy luminosity function 
(e.g., Lilly et al.\ 1995; Wolf et al.\ 2003; Willmer et al.\ 2005) 
have to be interpreted in terms of an unchanging characteristic stellar mass $M^*$ and a rapidly evolving mean stellar $M/L$ ratio (and 
therefore a rapidly evolving average color with redshift, 
as is indeed observed).  

Focusing instead on the red sequence (diamonds and solid lines), 
there is rapid evolution in their stellar mass function.  Assuming
a constant faint-end slope (which we cannot meaningfully 
constrain in this work\footnote{Kodama et al.\ (2004) and
de Lucia et al.\ (2004) present deeper data which show some
evidence for a deficit of faint
red-sequence galaxies at $z \sim 1$ compared to the present-day
red-sequence galaxy stellar mass function.}), the formal 
fits prefer evolution in $\phi^*$ (such that $\phi^*$ increases 
towards the present day) without significant change in $M^*$.
It is worth noting that this result is insensitive to the detailed choice
of color--mass cut used to separate the red and blue populations ---
as long as the cut is somewhere in the `gap' between the red and blue 
populations.

\subsection{The evolution of stellar mass densities}

\begin{figure}
  \centering
  \includegraphics[width=9.1cm]{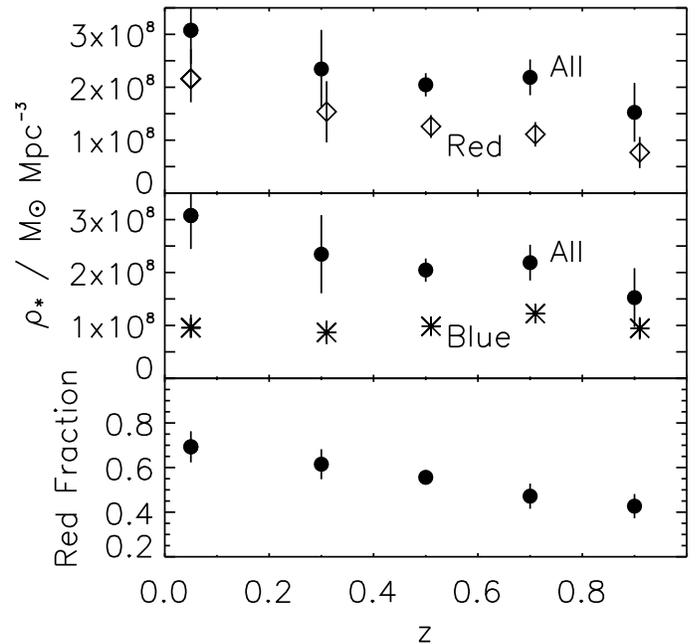}
\caption{The integrated stellar mass density as a function of
redshift.  In the upper two panels the total mass density for all galaxies (filled circles) is compared with those for red-sequence galaxies (diamonds), and for blue cloud galaxies (asterisks) seperately.  The lower panel shows
the fraction of mass in red sequence galaxies 
as a function of redshift. In all cases, mass functions
are integrated down to zero mass and error bars come from
field-to-field variation divided by $\sqrt{2}$.  
The $z = 0$ datapoint is taken from Bell et al.\ (2003).  }
\label{rho_z}
\end{figure}

The average stellar mass density $\rho_* \equiv \int{\Phi(m) dm}$
is calculated by integrating the Schechter functions of Table
\ref{Schechter} between $m = 0$ and $\infty$. The mass density
$\rho_*$ for {\it all} galaxies is calculated for all
three COMBO-17 fields; the larger of {\it i)} 
the field-to-field scatter, divided
by $\sqrt{2}$, or {\it ii)} the predicted variance, following
Somerville et al.\ (2004) given a number density and redshift, 
is adopted as the cosmic variance 
error bar.  The results in 
Fig. \,\ref{rho_z} show, that the total stellar mass has increased 
substantially between $z \sim 1$ and the present day.  Interestingly,
much of the evidence for this growth comes from the comparison
to the local stellar mass density\footnote{In this context, it is 
worth noting that the local total stellar mass densities, and indeed those
split by color, calculated
by a variety of different groups are very similar with $<0.1$\,dex
scatter if IMFs and cosmologies are accounted for, giving confidence
in the robustness of the locally-measured stellar mass density
estimates (e.g., Cole et al.\ 2001; Kochanek et al.\ 2001; Bell et al.\ 2003;
Baldry et al.\ 2004; Panter et al.\ 2004).}.  
Importantly, this stellar mass growth 
is dominated by the growth of the total stellar mass in red-sequence
galaxies, while the total stellar mass in blue cloud galaxies is 
roughly constant over the last 8\,Gyr.

\section{Discussion}
\label{discussion}

The COMBO-17 data and analysis constitute one of the first opportunities 
to combine
large samples of galaxies to z$\sim 1$, 
rather precise redshifts (average $\sigma_z / (1+z) \simeq 0.02$), 
and accurate rest-frame colors (U-V)$_{\rm rest}$ luminosities 
and stellar mass estimates.
In particular, its rest-frame colors are accurate enough 
to allow for the first time the
exploration of the mass growth of red and blue galaxies 
separately.

\begin{figure}
  \centering
  \includegraphics[width=9.1cm]{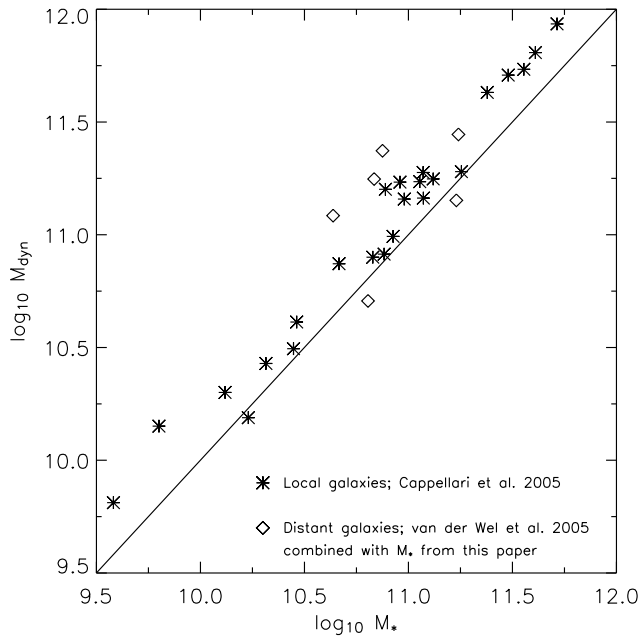}
\caption{A comparison 
of stellar and dynamical masses.  Accurate 
stellar and dynamical masses (asterisks) are taken 
from Cappellari et al.\ (2005) for a sample 
of local early-type galaxies.  Dynamical masses
(van der Wel et al.\ 2005) for 6 COMBO-17 galaxies
at $0.6 < z < 1.05$ with robust stellar masses are shown as diamonds.   }
\label{dynmass}
\end{figure}

\subsection{SED-based Stellar Mass Estimates}

Through extensive SED matching, based on the PEGASE
models (Fioc and Rocca-Volmerange 1997), we have illustrated the strength
and limitations of stellar mass estimates based on this approach. We have shown that for a given IMF (e.g. Kroupa et al.\ 1993), the present modelling indicates that SED-based
$M/L$ estimates are better than a factor of two for most individual galaxies,
with systematic sample uncertainties probably $< 30\%$ ({\it c.f.} Fig.\,\ref{ML_BR} and Section 3.5, see also 
Drory et al.\ 2004b).  We stress,
however, that the masses derived here simply scale for 
different IMFs (below 2M$_\odot$);
e.g. for a Salpeter (1955) IMF all masses would be higher by a factor of 
1.8.

Note that the analysis of the redshift dependence of stellar masses
does not require an accurate absolute calibration of the 
stellar mass scale; rather, relative consistency between the stellar mass
scale at $z \sim 1$ and $z \sim 0$ is sufficient.
Figure 11 shows 
the comparison between a sample of local early-type
galaxies with accurately-derived stellar 
and dynamical masses (Cappellari et al.\ 2005; a Kroupa 
2001 IMF is adopted in that work, which has a very similar mass
scale to the Kroupa et al.\ 1993 IMF) and 
a sample of 6 distant early-type galaxies in the 
Chandra Deep Field South with robust dynamical mass
estimates from deep spectroscopy (van der Wel et al.\ 2005) 
and stellar mass estimates from this paper\footnote{A substantial
fraction of the van der Wel et al.\ sample were at or below
the COMBO-17 $m_R < 24$ limit and had poorly-estimated
or undefined redshift.}.  There is a clear
relationship between stellar and dynamical mass\footnote{Note that 
dynamical mass should always be in excess of stellar mass 
(allowing the existence of dark matter in the inner parts of galaxies);
clearly a Kroupa et al.\ (1993), Kroupa (2001) or Chabrier (2003)
IMF satisfies these constraints whereas the Salpeter (1955) IMF
(which produces stellar masses $\sim 0.25$\,dex larger) would
violate this constraint (see Cappellari et al.\ 2005; Bell \& de Jong 
2001).}, and this relationship
seems independent of redshift at least out to $z \sim 1$.  This 
gives us some confidence that stellar population masses 
are being estimated consistently for local and distant samples, 
thus evolutionary trends between COMBO-17 and local samples should 
be relatively robust.

Our study illustrates quite dramatically how galaxy samples,
even when selected at red observed bands, are very far from a 
stellar mass limited sample.
The difference in stellar mass between the
bluest and reddest galaxies at a given flux limit is a factor 8 at 
$z \simeq 0.3$ and more than 30 at $z \simeq 1$ (where 
the sample-selecting R-band corresponds to 
$\lambda_{\rm rest}\sim 360$\,nm).

\subsection{The relationship between color and stellar mass}

Our most conspicuous  astrophysical result 
is that at $z < 1$, the majority of the
most massive galaxies (M$_* > 10^{11}$M$_\odot$) are 
on the red sequence (see Figs.\ 7 and 9). This
has qualitatively been clear from color--luminosity diagrams
(e.g., Bell et al.\ 2004), but Figs.\ 7 and 9
demonstrate that it holds true for the top decade in stellar mass.
It is not clear, at this stage, what this result implies in terms
of galaxy evolution.  The stellar mass functions (Fig.\,9) show a very modest increase in the number 
of red massive (M$_* > 10^{11}$M$_\odot$) galaxies and a  
modest decrease in the number of  blue massive ones 
from $z = 1$ to the present day.  Yet, it 
is important to bear in mind that uncertainties in stellar masses
and photometric redshifts have a particularly strong effect on the 
high-mass end of stellar mass functions; therefore, we argue that 
it is not clear whether the massive galaxy population was largely 
in place and non-star-forming at $z \sim 1$ (e.g. Saracco et al.\ 2005), 
or whether there 
has been some growth of massive non-star-forming galaxies
through processes such as dry mergers (e.g. Bell et al.\ 2005b).

Over the last decade, there has been extensive 
evidence for what has become known as 'downsizing':
while massive galaxies form stars at high redshift (e.g., Chapman et al.\
2003; Daddi et al.\ 2005), only low-mass galaxies form
significant numbers of stars at the present  (Cowie et al.\ 1996; 
Heavens et al.\ 2004; 
Juneau et al.\ 2004; Bauer et al.\ 2005;
Bell et al.\ 2005a; Tanaka et al.\ 2005; Kelm et al. 2005).  
The color-split mass functions presented in this paper allow 
us to explore some aspects of this effect.  Figure 9, for
instance, shows the evolution of 
the mass at which the red sequence ($=$ non-star-forming)
and blue cloud ($=$ star forming) stellar mass functions 
cross\footnote{This definition is identical to Bundy et al.'s (2006) transition mass $M_{tr}$, and is} very similar to that adopted by
Kauffmann et al.\ (2003), who explore the distribution of 4000\,{\AA}
break strengths as a function of stellar mass.: above this mass
the majority of galaxies is red and below it
the majority of galaxies is blue.
We find that this characteristic mass is 
$\sim 10^{10.8} M_{\odot}$ at $z = 1$ (in agreement with estimates presented in Bundy et al.\ 2006) and 
decreases by a factor of 5 or more to $\sim 10^{10} M_{\odot}$ at 
$z \simeq 0$.  Yet, the physical significance of this `transition mass' is 
not obvious as the stellar mass function of blue 
galaxies does not change significantly
in the epoch $0<z<1$, suggesting that the maximum mass that galaxies 
reach while forming stars remains roughly the same.
The apparent evolution of the 'transition mass' is driven 
entirely by the increasing prominence of the non-star-forming
red sequence galaxies.  This emphasizes the need for 
care when discussing `downsizing', and the utility of large
samples of galaxies with robust
estimates of stellar masses for studying the physics
driving the order-of-magnitude decrease in cosmic
star formation rate since $z \sim 1$ (e.g., Le Floc'h et al.\ 2005).

\subsection{The build-up of stellar mass}

It is interesting to compare the COMBO-17-derived 
stellar mass density evolution since $z \sim 1$ with 
other work that has studied the build-up
of the galaxy population since $z \sim 4$.  
The results are shown in Fig.\ 12.  The top panel shows the 
total stellar mass in galaxies as a function of redshift 
(integrated to zero mass) for a variety of studies.
The stellar masses have been adjusted to be consistent with
a Kroupa et al.\ (1993), Kroupa (2001) or Chabrier (2003) IMF;
all three IMFs have a stellar $M/L$ at a given optical
color that are 0.25 dex lower than for a Salpeter (1955) IMF, 
and all three IMFs are consistent with dynamical constraints 
(Bell \& de Jong 2001; Cappellari et al.\ 2005), while a Salpeter
IMF over-predicts the dynamical masses of many galaxies.
Despite the inevitable uncertainties associated with 
estimating stellar masses, and cosmic variance, Fig.\,12 illustrates that the different studies have come to a reasonably consistent picture of the growth 
of stellar mass with cosmic time.

\begin{figure}
  \resizebox{\hsize}{!}{\includegraphics{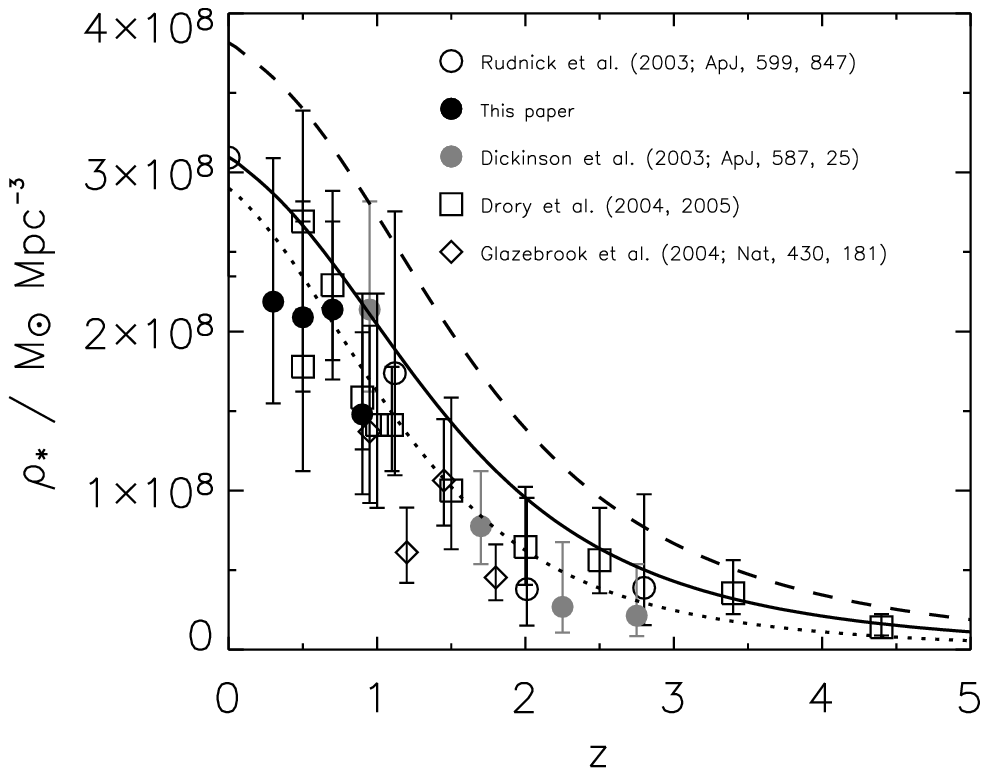}}
  \resizebox{\hsize}{!}{\includegraphics{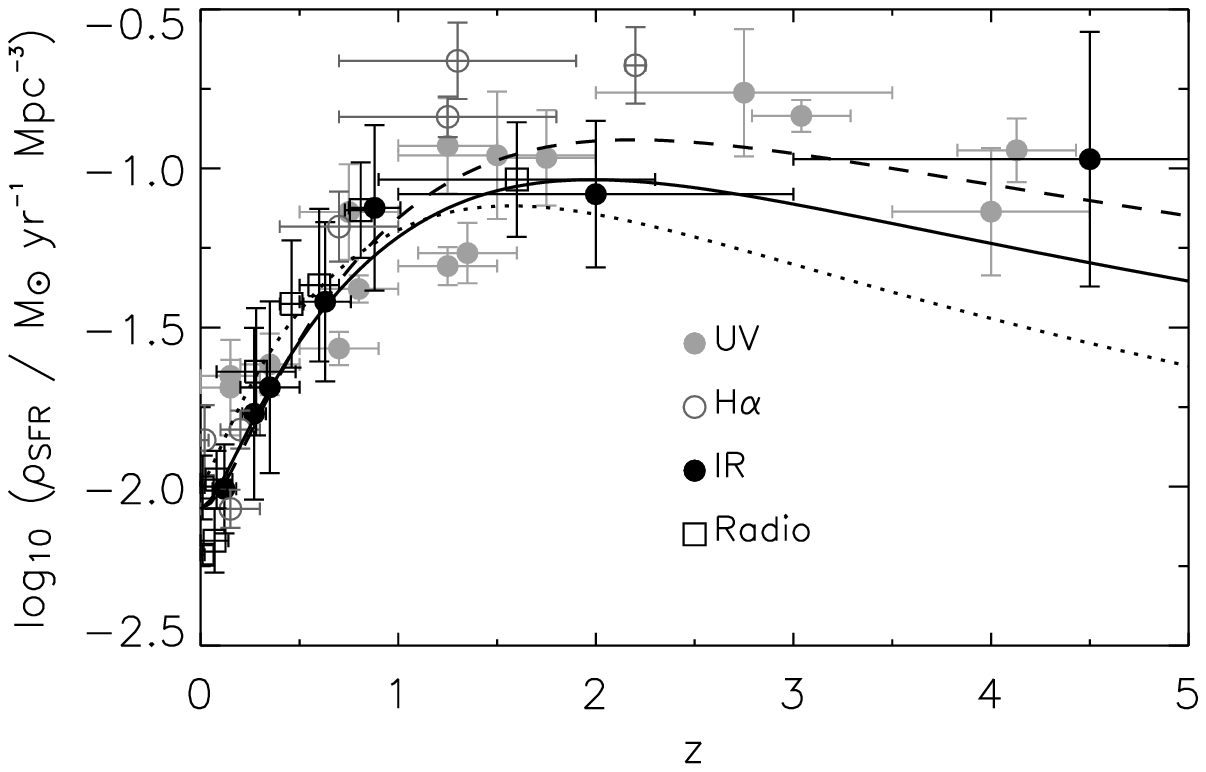}}
 \caption{Top: the growth of stellar mass density with 
cosmic time, as estimated from a number of surveys.  Bottom:
the evolution of the cosmic star formation rate, adapted
from Hopkins (2004).  In both panels, the stellar IMF has been 
adjusted to be consistent with Chabrier (2003); a Kroupa
et al.\ (1993) IMF has a stellar $M/L$ consistent with such an 
IMF but would produce star formation rates a factor of three
higher than the Chabrier (2003) IMF.  The lines show
different assumptions for the evolution of the cosmic
star formation history and the associated growth of 
cosmic stellar mass: the dotted line shows the fit optimized
to reproduce the growth of stellar mass, the dashed line
fits the star formation rate better, and the solid line is a 
reasonable compromise between the two.}
  \label{SFhistory}
\end{figure}

This growth of stellar mass should be consistent with the integral
of current measurements of the cosmic SFR.  The lower
panel of Fig.\ 12 shows the evolution of the cosmic SFR (adapted from
Hopkins 2004).  We have chosen not to convert the cosmic SFR measurements
from a Salpeter (1955) IMF to a Kroupa et al. (1993) IMF in this work: as
the power-law index of a Kroupa et al. (1993) IMF is substantially steeper
than those of a Salpeter, Chabrier or Kroupa (2001) IMF for high-mass
stars (which dominate the SFR measurements, but do not significantly
affect broad-band SED-derived stellar masses), the use of a Kroupa et al.  
(1993) stellar IMF in what follows would lead to a factor of three
over-production of stars.  Instead, we adopt a Chabrier (2003) IMF (which
has the same high-mass IMF slope as a Salpeter IMF and very similar
stellar M/Ls to a Kroupa et al. (1993) or Kroupa (2001) IMF), and
ask the question whether the integral of the cosmic star formation rate 
is consistent with the observed growth of stellar mass.  We show
three cosmic star formation histories and their integrals (accounting
fully for gas recycling).  The dashed line is a fit optimized
to reproduce the evolution of cosmic star formation rate: this 
tends to produce a little too much stellar mass at all redshifts.
The dotted line provides a much better fit to the build-up of stellar
mass, but appears to undershoot somewhat the star formation rate
estimates at $z > 1$.  The solid line is a reasonable compromise 
between matching the cosmic star formation rate and build-up of 
stellar mass.  Despite these interesting tensions at the 30--50 \% level --- which may indicate
deficiencies in stellar mass or SFR estimates ({\it e.g.} due to a globally incorrect IMF and/or one which varies from galaxy to galaxy), or may
betray inappropriate extrapolation to total densities from the observed
galaxies  --- it is nonetheless fair to say that 
the match between cosmic SFR and the build-up of 
stellar mass is rather good.

The relatively precise rest-frame colors of COMBO-17 allowed us to separate
out the evolution of  $\langle \rho_*\rangle (z)$ in red and blue galaxies.
While a similar split has not yet
been carried out for estimates of cosmic star formation rate, it
is interesting to note that study of the UV (Wolf et al.\ 2005)
and UV$+$IR (Bell et al.\ 2005a) luminosity of galaxies has demonstrated
that the bulk of star formation takes place in spiral galaxies residing in the blue cloud.
Thus, one would naively expect that the growth of stellar
mass at $z < 1$ would be dominated by a growth in the total stellar
mass in blue galaxies.  Yet, Fig.\ 10 shows that the total 
stellar mass growth takes place mainly in the red sequence.  
Thus, stars form in blue galaxies, but a substantial fraction 
of the star-forming galaxy population 
must globally truncate their star formation and fade
and redden onto the red sequence at $z < 1$.  A discussion 
of this point is deferred to a future paper.

\section{Summary and Conclusions}

We have presented estimates of the stellar masses, derived from SED-fitting
of the COMBO-17 flux-points using PEGASE population synthesis models,
for 25\,000 galaxies to $z = 1$, with known redshifts, rest-frame colors and
luminosities. From this data set we were able to show the following.

\begin{itemize}
\item Since $z\sim 1$ the majority of galaxies with $M>10^{11} M_{\odot}$ 
have been on the red-sequence, i.e., have 
essentially completed the build-up of their stellar mass.

\item As a consequence, the stellar mass function of red and blue galaxies 
differ strongly from each other at all redshifts from 0 to 1. Red galaxies always dominate the massive end, while blue galaxies dominate the faint end.  
The mass at which the mass functions cross decreases from 
$M_* \sim 10^{10.8} M_{\odot}$ at $z = 1$ to
$\sim 10^{10} M_{\odot}$ at the present day.  This evolution in the 
`transition mass', above which most galaxies are not
forming stars, is driven primarily by the increasing prominence
of the non-star-forming galaxy population at later times.

\item The total stellar mass in galaxies has roughly doubled
in the 8 Gyr since $z = 1$.  Assuming a Chabrier (2003) IMF, 
the stellar mass evolution and star formation rate evolution 
are reasonably consistent: a Kroupa et al.\ (1993) IMF
(which has a stellar $M/L$ for a given SED which is very similar
to those of a Chabrier IMF) has a relatively low number of 
high-mass stars and no consistent interpretation of the 
cosmic star formation rate and growth of stellar mass 
can be found.

\item Somewhat counter-intuitively, the stellar mass 
function of blue, star-forming galaxies 
has stayed more or less constant since
$z = 1$; 
the mass in ``dead'' red-sequence galaxies has almost tripled over that period.
While at $z = 1$ half the stars lived in 
red sequence galaxies and half in star-forming
galaxies, 2/3 of all stars at the present epoch live in red-sequence galaxies.

\item Even with 25\,000 galaxies in three disjoint 
fields of $30^{\prime}\times30^{\prime}$
each, field-to-field variations limit the precision of $\langle \rho_*\rangle (z)$ estimates
to $\sim 30\%$.

\end{itemize}

There are a number of obvious steps which need to be taken to 
improve the accuracy and extend the scope of the work presented in this paper.
Larger areas, more precise redshifts (particularly spectroscopic), 
and wider wavelength coverage will give increased precision on 
the form of the mass function (especially at the high-mass end; critical
for exploring the influence of galaxy mergers on the stellar mass function) 
and increased redshift range.  
Larger samples of dynamical masses and improved templates
will increase the accuracy (i.e., reduce systematic error) 
in the mass functions.  Finally, accurate determinations of 
star formation rates from Spitzer/Herschel/ALMA 
will allow star formation and the build-up of stellar
mass to be intercompared at a much more detailed and informative
level than is currently possible, further elucidating the role
of {\it in situ} star formation vs.\ galaxy mergers in driving the
mass evolution of the galaxy population.

\begin{acknowledgement}
We are grateful to an anonymous referee who made numerous suggestions
how to improve this paper.
We thank Roelof de Jong and Rachel Somerville for interesting discussions
which helped to shape this work, and Michele Cappellari for the data for Fig.\ 11 in electronic form.
E.\ F.\ B.\ was supported by the European Community's Human
Potential Program under contract HPRN-CT-2002-00316 (SISCO).

\end{acknowledgement}

\end{document}